\documentclass[11pt,a4paper]{article}
\pdfoutput=1

\usepackage{jcappub}

\usepackage{amsmath,amsthm,amssymb,amsfonts}

\usepackage{dcolumn}
\usepackage{bm}
\usepackage[latin1]{inputenc}
\usepackage[spanish,english]{babel}
\usepackage{amsfonts}
\usepackage{amssymb}
\usepackage{graphicx}
\usepackage{verbatim}
\usepackage{color}

\newcommand{\be}{\begin{equation}}
\newcommand{\ee}{\end{equation}}
\newcommand{\bea}{\begin{eqnarray}}
\newcommand{\eea}{\end{eqnarray}}

\begin{document}

\title{Enhanced local-type inflationary trispectrum from a non-vacuum initial state 
}

\author[a]{Ivan Agullo,}\emailAdd{agullo@gravity.psu.edu}
\affiliation[a]{ {\footnotesize  Institute for Gravitation and the Cosmos, The Pennsylvania State University, University Park, PA 16802, USA}}
\author[b,c]{Jose Navarro-Salas,}\emailAdd{jnavarro@ific.uv.es}
\affiliation[b]{ {\footnotesize Departamento de Fisica Teorica and
IFIC, Centro Mixto Universidad de Valencia-CSIC.
    Facultad de Fisica, Universidad de Valencia,
        Burjassot, Valencia 46100, Spain,}}\affiliation[c]{ {\footnotesize Physics Department, University of
Wisconsin-Milwaukee, P.O.Box 413, Milwaukee, WI 53201, USA}}

\author[c]{and Leonard Parker}\emailAdd{leonard@uwm.edu}

\date{December 2, 2011}

\abstract{
We compute the primordial trispectrum for curvature perturbations produced during cosmic inflation in models with standard kinetic terms, when the initial quantum state is not necessarily the vacuum state. The presence of initial perturbations  
enhances the trispectrum amplitude for configuration in which one of the momenta, say $k_3$, is much smaller than the others,  $k_3 \ll k_{1,2,4}$. For those squeezed configurations 
the trispectrum acquires the so-called local form, with a scale dependent amplitude  that can get values of order
$ \epsilon \left ({k_1}/{k_3} \right)^2$.
This amplitude can be larger than the prediction  of the so-called Maldacena consistency relation
by a factor $10^6$, and can reach the sensitivity of forthcoming 
observations, 
even for single-field inflationary models. }

\maketitle

\section{Introduction} \label{introduction}

The inflationary paradigm provides a compelling argument to account for the origin of the cosmic inhomogeneities that we observe in the CMB and galaxy distribution. Primordial inhomogeneities arise naturally during inflation as the result of the amplification of quantum fluctuation by the exponential expansion of the universe. The detailed observations obtained by the WMAP satellite \cite{wmap7} indicate that the temperature fluctuations of the CMB follow, with high precision, a Gaussian statistical distribution with an almost scale invariant spectrum. Those observations are compatible, in a non trivial way, with the predictions of inflation, even for the minimal implementation based on a single scalar field slowly rolling down the potential. 

The minimal implementation of inflation has been generalized in different directions. The effective theory of inflation \cite{cheung,weinberg2008} embeds in an elegant way generic extensions  
to models with multiple scalar fields or non-standard kinetic terms with higher derivatives. An important question is then, can observations distinguish between those different extensions? The predictions regarding the Gaussian part of the spectrum of inhomogeneities arising from inflation are highly degenerate for different models. In other words, the power spectrum alone has limited potential in revealing detailed information about the dynamics of inflation.  However, the deviations from a perfect Gaussian distribution, the so-called non-Gaussianities, have been proven to contain valuable information to distinguish among different models of inflation. Nowadays, the observation of non-Gaussianities in the CMB and large scale structure is considered a sharp tool to extract information about the physics of the field or fields driving inflation \cite{whitepaper}. 

In this paper we address the consequences of extending the minimal model by generalizing the initial conditions. The computation of the spectrum of non-uniformities created during inflation requires a basic assumption concerning the quantum state describing scalar metric perturbations at the onset of inflation. Such a state is usually chosen as a natural extension of the Bunch-Davies vacuum \cite{bunch-davies} of the exact de Sitter space, to the quasi-de Sitter inflationary spacetime.  The assumption of  the exact  vacuum state may seem, however,  too restrictive. An 
often stated 
viewpoint is that the huge inflationary expansion will dilute any quanta present in the initial state, washing away any deviation from the vacuum. However, this intuition is not completely accurate in the quantum theory. Quantum field theory in curved backgrounds predicts that the presence of quantum perturbations at the onset of inflation will generically produce {\em stimulated} creation of additional quanta due to the spacetime expansion \cite{parker66,parker69}. The stimulated creation of quanta 
compensates for the dilution during inflation, making the deviation from a vacuum state potentially observable in the present universe \cite{agullo-parker}.

Two relevant question arise then; can cosmological observations {\em in the present universe} be sensitive to the details of the {\em initial conditions} of the effective field theory of inflation? And the second important question, can observations {\em disentangle} the effects of the initial conditions from other aspects of the effective theory? The first question has been addressed by different authors with a generic positive answer, under quite reasonable assumptions \cite{gasperiniveneziano,bhattacharya,boyanovsky, holman-tolley, meerburg, chen0, magueijo, holman, paban, ganc, agullo-parker, chialva, paban2, kundu}. In this paper we address those questions by analyzing the inflationary trispectrum. \\

It was pointed out in \cite{chen0, holman-tolley} that, as already happens for other aspects of the effective theory of inflation, non-Gaussianities are sensitive to the initial conditions. It was shown there that the bispectrum (the three-point function in momentum space) for curvature perturbation shows a considerable enhancement when the three momenta involved in the bispectrum form a flattened or folded triangle ($k_1\approx k_2\approx  k_3/2$), and the state describing metric perturbation at the onset of inflation departs from the vacuum state.
(Throughout this paper, 
we use the notation $k\equiv |\vec{k}|$.) More recently, it was shown in \cite{agullo-parker} that
when a non-vacuum state is considered, the bispectrum shows an important and unexpected enhancement for 
the configuration in which the momenta form
a {\em squeezed} triangle ($k_1\approx k_2\gg k_3$). In that squeezed momentum configuration, the primordial bispectrum acquires the so-called local form 
\be \label{bi} B_{\zeta} (k_1,k_2,k_3)=   \frac{12}{5} \, f_{NL}\,   P_{\zeta}(k_1)P_{\zeta}(k_3) \ , \ee
with a momentum dependent parameter $f_{NL}= 5/3 \, \epsilon\, (k_1/k_3)\, \gamma$, where $\gamma=\mathcal{O}(1)$ is a factor depending on the characteristics of the initial state, and $\epsilon$ is the standard slow-roll parameter (see \cite{agullo-parker} for details). When compared to the vacuum state prediction \cite{maldacena}, this $f_{NL}$ parameter contains an extra factor $k_1/k_3\gg 1$, that can be as large as several hundred for the range of momenta accessible in observations.

The goal of the present paper is to deepen the analysis of the effects of an excited initial state by studying the inflationary trispectrum of curvature perturbations (the four-point function in momentum space). In section \ref{quantum state}, we briefly discuss physical criteria for constraining the initial quantum state. Those constraints come from theoretical arguments as well as from present CMB observations. In section \ref{section trispectrum}, we work out the primordial trispectrum for curvature perturbations, by generalizing the computations in \cite{seery-lidsey-sloth,seery-sloth-vernizzi} to the case of a non-vacuum initial state. Our computations are explicitly carried out for a single-field model with canonical kinetic term, but are easily generalized to multi-field models with flat field-space metric. We write our results for initial conditions given by a pure quantum state resulting from a Bogolubov transformation of the 
Bunch-Davies
 vacuum, although a more general initial state would leave our main conclusions unchanged. We find that the presence of initial quanta 
 greatly
 enhances the primordial trispectrum  in the squeezed limit, in which one of the momenta is much smaller that the other three. In this limit, the trispectrum acquires the so-called local form, with a scale dependent amplitude that can reach the sensitivity of forthcoming observations. We discuss, in section \ref{discussion and conclusions}, the consequences of our results for the so-called 
 Maldacena  consistency relation \cite{maldacena, creminelli-zaldarriaga,chen}, 
 and on the related possibility of 
 disfavoring
 slow-roll inflation by observing local-type non-Gaussianities. Throughout this paper we use the standard convention $c=\hbar=1$,  and $M_P=1/\sqrt{8 \pi G}$, with $G$ the Newton constant.

\section{Initial quantum state} \label{quantum state}

The computation of the spectrum of inhomogeneities generated during inflation requires a basic assumption about the quantum state describing those perturbations at the onset of inflation. For instance, in a universe in which inflation starts at a certain finite time after  
a quantum gravity era replaces the Big Bang singularity,
the state of perturbations at the onset of inflation may deviate from the vacuum state as a consequence of a non-trivial pre-inflationary evolution. 
In such a scenario,
the pre-inflationary evolution could produce a state for 
the perturbations that has
a non-zero content of quanta as compared to the vacuum state, 
as well as containing non-Gaussianities. Therefore, the most general state for perturbations at the onset of inflation  would be a non-vacuum, non-Gaussian state.
 That state could also be a mixed state, which 
 can
 naturally arise in the context of inflation in which the observable universe results from the enormous expansion of a small patch in a much larger universe. The state of such a patch would naturally be a mixed state, even if the global state of the larger universe were a pure state, because many features of the global state would 
not
 be accessible to our observable universe \cite{vonneumann}. 
For brevity, we obtain our results for the case of a pure state that differs from the vacuum. However,
it is straightforward to extend our discussion to mixed states, as in \cite{agullo-parker}.

A reasonable assumption about primordial non-Gaussianities is that, 
as a consequence of the rapid and large
exponential expansion, 
the non-Gaussianities generated during inflation will generically dominate the primordial spectrum, and any potential deviations from Gaussianity created before inflation can be consistently neglected. It is then natural to restrict to a Gaussian state at the onset of inflation. 
It is easy to show that the most general pure, Gaussian state can be described by Bogolubov transformation
 of the vacuum state \cite{branustein-pati}. Those states include, for instance, coherent and squeezed states.
We will then restrict our analysis below to states at the onset of inflation given by Bogolubov transformation of the Bunch-Davies vacuum for scalar perturbations.
(A Bogolubov transformation of a vacuum state for scalar particles was first used in the cosmological context in \cite{parker-fulling}, to show that quantum effects can avoid the big bang singularity.)

The initial state can be specified by expanding the field operator of the primordial curvature perturbation $\zeta$ \cite{dodelson, lyth-liddle, weinbergbook} in Fourier modes
\be \hat{\zeta}(\vec{x},\tau)=\int \frac{d^3 k}{(2\pi)^3} \ \hat{ \zeta}_{\vec{k}}(\tau) \ e^{i \vec{k}\vec{x}}\ ,   \hspace{1cm}  \hat{\zeta}_{\vec{k}}(\tau)=A_{\vec{k}} \, \bar{\zeta}_{k}(\tau)+ A_{-\vec{k}}^{\dagger} \, \bar{\zeta}_{k}^*(\tau) \ ,\ee 
where $\tau$ indicates the conformal time. The general mode functions 
for de Sitter inflation 
can be written as a linear combination
$\bar{\zeta}_{k}(\tau)=\alpha_k \ {\zeta}_{k}(\tau)+\beta_k \ {\zeta}_{k}^*(\tau)$, where $|\alpha_k|^2-|\beta_k|^2=1$, 
and
\be \label{modes}{\zeta}_k(\tau)= \frac{H^2}{\dot \phi_0} \frac{(1+ik \tau)}{\sqrt{2 k^3}}e^{-ik\tau} \ , \ee
are de Sitter invariant modes.\footnote{If we slightly depart from the pure de Sitter geometry, as in slow-roll inflation, the above modes generalize to Hankel functions (see, for instance, \cite{weinbergbook}).} In the above expression, $\phi_0(t)$ is the homogeneous part of the inflaton field sourcing the inflationary background expansion, and $H\equiv\dot a/a$ is the Hubble function, with $a(t)$ the scale factor and the dot indicating derivative with respect to the cosmic time. With the normalization (\ref{modes}), the creation and annihilation operators $A_{\vec{k}}$ and $A_{\vec{k}}^\dagger$ satisfy the 
commutation
relations $[A_{\vec{k}},A_{\vec{k'}}^\dagger]=(2\pi)^3\delta^3(\vec{k}-\vec{k'})$. In the limit 
in which $\alpha_k=1$ and $\beta_k=0$ for all $k$, the state annihilated by all the operators $A_{\vec{k}}$ is the so-called Bunch-Davies vacuum state. For arbitrary values of the Bogolubov coefficients $\alpha_k$ and $\beta_k$ 
(with $|\alpha_k|^2-|\beta_k|^2=1$)
we 
get a Bogolubov transformation of the Bunch-Davies vacuum. Such a state can be interpreted as containing an average number density of quanta per unit proper volume
$(2\pi a)^{-3}|\beta_k|^2d^3k$ with momenta near $\vec{k}$ in the range $d^3k$, as compared to the Bunch-Davies vacuum.

There are some restrictions on the ultraviolet and infrared behavior of the coefficients $\alpha_k$ and $\beta_k$.
Regarding the ultraviolet (UV) behavior, the adiabatic condition \cite{parker-toms}, or similarly the Hadamard condition, requires that $\alpha_k \to 1$ and $\beta_k \to 0$ sufficiently fast as $k \to \infty$. This condition ensures that the UV divergences appearing in expectation values of relevant quantum operators can be systematically cured by methods of renormalization and regularization.

In addition, the behavior of $\alpha_k$ and $\beta_k$ as $k \to 0$ may also be restricted to ensure the absence of infrared (IR) singularities in the case where spatial slices of constant cosmic time have the $\mathbb{R}^3$ topology.
A requirement for the initial state is that the two-point function,
\bea  \label{integ} \langle {\hat \zeta}(\vec{x}_1, \tau){\hat \zeta}(\vec{x}_2, \tau)\rangle &=&  \frac{H^4}{2\dot \phi^2_0} \int_0^{\infty} \frac{dk}{k} \Big[(1+2|\beta_k|^2)  |(1+ik\tau)|^2  +2\  {\rm Re}\big[\alpha_k\beta_k^{*}(1+ik\tau)^2e^{2ik\tau})\big]\Big]    \nonumber  \\ &\times& \frac{\sin (k|\vec{x}_1-\vec{x}_2|)}{k|\vec{x}_1-\vec{x}_2|} \ , \eea
needs to be non-singular. In the limit $k\to 0$ the integrand above behaves as  $|\alpha_k+\beta_k|^2 k^{-1}$. The absence of IR divergences in (\ref{integ}) requires then $|\alpha_k+\beta_k|^2 \to 0$, as $k \to 0$ \cite{ford-parker}.\footnote{One has also ultraviolet divergences when $\vec{x}_1\to\vec{x}_2$ due to integration for large $k$. These  divergences can be cured by renormalization in an expanding universe, as has been studied in \cite{grg09}. The (momentum-space) subtracting terms required to cancel out these divergences can potentially lead to modifications of the power spectrum at scales relevant for cosmology. Because the potential effects of renormalization do not interfere with the effects arising from a non-vacuum state, we shall not consider renormalization in this paper.}   This conditions imposes restriction in both, the norm of $\beta_k$ and the relative phase, $\theta_k$,  between $\alpha_k$ and $\beta_k$. On the one hand, it requires $N_k\equiv|\beta_k|^2\neq 0$ as $k\to0$. This excludes, in particular,  an exact Bunch-Davies state with $\alpha_k =1$ and $\beta_k= 0$ for all $k$ (in other words, the Bunch-Davies vacuum is not a Hadamard state \cite{waldbook}\footnote{This is equivalent to the statement  \cite{allen} that it is not possible to construct a (Hadamard) quantum state invariant under the full de Sitter group for a massless scalar field.}). This is also true in slow-roll inflation where the modes (\ref{modes}) are replaced by Hankel functions.  On the other hand, the relative phase $\theta_k$ is required to behave as $\cos \theta_k \sim (1+2N_k)/(2\sqrt{N_k(N_k+1)})$ as $k\to 0$. For instance, this conditions forbids $\theta_k \propto k$, but allows $\theta_k \approx$ const, in the limit $k \to  0$.
 
On the other hand, as shown in \cite{lyth, boyanovsky, anderson}, by imposing negligible back reaction on the inflationary expansion, the occupation number $N_k$ in a mode with  momentum ${\vec k}$ is constrained to be not much larger than 1, although this constraint may be stronger in case the duration of inflation is much larger that the usual assumption of around $65$ e-folds. Additional constraints on the initial state come from observations \cite{wmap7} of the amplitude of the power spectrum and spectral index, $n_s$, characterizing its dependence on $k$.  However, the power spectrum alone has limited potential in revealing detailed information about the initial state, as essentially it only  
constrains
the quantity $\left| \frac{d\ln{ |\alpha_k+\beta_k|}}{d \ln k}\right| $ to be not much bigger than the slow-roll parameters $\epsilon \sim |\delta|\sim 1/100$, defined in terms of the Hubble rate as $\epsilon \equiv -\dot{H}/H^2$ and $\delta \equiv \ddot{H}/(2\dot{H} H)$ \cite{agullo-parker}.

\section{The trispectrum for an excited initial state} \label{section trispectrum}

This section is 
devoted
to the computation of the primordial trispectrum for the curvature perturbations $\zeta(\vec{x},\tau)$ generated during inflation when the initial state is allowed to deviate from the standard vacuum state. For concreteness in the computation, in this section we consider an initially pure, Gaussian state given by a Bogolubov transformation of the vacuum.  
We specify the initial state at a time $\tau_0$, a few e-foldings before the smallest momentum $\vec{k}$ involved in the trispectrum exits the Hubble sphere during inflation.

Our computations proceed by generalizing the detailed analysis in \cite{seery-lidsey-sloth,seery-sloth-vernizzi}, where the trispectrum of the primordial curvature perturbations was computed for an initial Bunch-Davies vacuum state. We will not repeat here all the details of the computation, and the reader is referred to \cite{seery-lidsey-sloth,seery-sloth-vernizzi} 
for further details. As in those papers, we explicitly write down the computations for a single-field inflationary model with canonical kinetic energy, although the main conclusions also apply to multi-field models with flat field-space metric.  

The primordial trispecturm, $T_{\zeta}$, is defined in terms of the four-point function in momentum space evaluated at late times during inflation $ \tau \gg \tau_{k_a}$, where $\tau_{k_a}$ is the Hubble exit time for the comoving momentum $k_a$, defined as the time at which $k_a/a(\tau_{k_a}) = H(\tau_{k_a})$,

\be \label{trispectrum}\langle \hat\zeta_{\vec{k}_1}\hat\zeta_{\vec{k}_2}\hat\zeta_{\vec{k}_3}\hat\zeta_{\vec{k}_4}\rangle= (2 \pi)^3  \delta(\sum_a \vec{k}_a) \ T_{\zeta}(\vec{k}_1,\vec{k}_2,\vec{k}_3,\vec{k}_4) \ . \ee
By convention, in the above expression only the connected part of the expectation value on the left hand side is considered. As shown in \cite{maldacena},  
by
virtue of the smallness of the slow-roll parameter $\epsilon$ during inflation, the result for the primordial correlation functions of curvature perturbations $\zeta$ is correctly reproduced by first computing the correlation function of inflaton perturbations $\delta \phi$ in the uniform curvature gauge, and by using the free solution 
for $\delta \phi$ in
de Sitter space.  
To leading order in slow-roll, the linear relation between modes $\zeta_k=(H/\dot{\phi}_0) \delta \phi_k$  allows 
us to obtain the correlation functions for $\zeta$ once the correlation functions for  $\delta \phi$ are known.

 If we restrict to the leading order contribution in slow-roll, 
the four-point  function for $\delta \phi_k$ has two contributions. The first contribution is given by the contact interaction of four scalar perturbations $\delta \phi_k$, and requires 
one 
to expand the action up to  
fourth
order in perturbations \cite{seery-lidsey-sloth}. The second contribution comes from the interaction mediated by the exchange of a tensor mode or graviton \cite{seery-sloth-vernizzi}. We will refer to those contributions as the contact interaction (CI) and the graviton exchange (GE) contributions, respectively. The action describing those interactions can be written as $S=S_{CI}+S_{GE}$, where \cite{seery-lidsey-sloth, maldacena}

\be \nonumber S_{CI}=\int d\tau d^3x \,  M_{P}^{-2}\, \left[ -\frac{1}{4} 	\theta_j \partial^2 \theta_j-a \delta \phi' \theta_j \partial_j \delta \phi-\frac{a^2}{4 H} \partial^{-2}\Sigma \left( (\partial_i \delta \phi)^2+ \delta\phi'^2\right) -\frac{3}{4} a^4 \partial^{-2} \Sigma  \partial^{-2} \Sigma\right] \ee

\be \nonumber S_{GE}=\frac{1}{2}  \int d\tau d^3x \ a^2 \gamma^{ij} \partial_i \delta\phi \partial_j \delta\phi \ , \ee
where $\gamma^{ij}$ represents tensor metric perturbations, and
 
\be  \theta_j= 2 a\, \partial^{-4} \left[\partial_m \partial_j \delta\phi' \partial_m \delta\phi+\partial_j \delta\phi' \partial^2\delta\phi-\partial^2 \delta\phi' \partial_j\delta\phi-\partial_m \delta\phi' \partial_m \partial_j \delta\phi \right] \, , \ee

\be \Sigma=\frac{1}{a} \left[ \partial_j \delta\phi' \partial_j \delta \phi+\delta\phi' \partial^2 \delta\phi \right] \, . \ee
We use a notation in which latin indices $i,j,l,m$ range form $1$ to $3$, the three dimensions of space, and latin indices $a,b,c,d$ range from $1$ to the number of momenta 
involved
in the correlation function ($4$ in the case of the trispectrum). By using the so-called ``in-in'' formalism \cite{weinberg05}, and after some effort, the
 contact interaction contribution 
 to the four-point function is given by

\bea \label{CI} \langle \hat{\delta \phi}_{\vec{k}_1}\hat{\delta \phi}_{\vec{k}_2}\hat{\delta \phi}_{\vec{k}_3}\hat{\delta \phi}_{\vec{k}_4}\rangle_{CI} =(2 \pi)^3 \delta(\sum_a \vec{k}_a)  \left(\prod_a \bar{\delta \phi}_{k_{a}}(0)\right) M_P^{-2} \, (M_{1234}+23 \ {\rm permut}) + \, {\rm c.c.} \ , \nonumber \eea
where

\be M_{abcd}=\left[ \frac{\vec{Z}_{ab}\cdot \vec{Z}_{cd}}{k_{ab}^6}+2 \, \frac{\vec{k}_b \cdot \vec{Z}_{cd}}{k_{cd}^4}+\frac{3}{4} \frac{\sigma_{ab} \sigma_{cd}}{k_{ab}^4}\right] I^{(1)}_{abcd}-\frac{1}{4 H} \frac{\sigma_{cd} }{k_{ab}^2} \left[\vec{k}_a \cdot \vec{k}_b \ I^{(2)}_{abcd}-I^{(3)}_{abcd}\right] \ ,\ee
with

\be \label{int1} I^{(1)}_{abcd}=-i \int_{\tau_0}^0 d\tau \ a^2(\tau) \  \bar{\delta \phi}'^{*}_{k_{a}}(\tau)\bar{\delta \phi}^*_{k_{b}}(\tau)\bar{\delta \phi}'^*_{k_{c}}(\tau)\bar{\delta \phi}^*_{k_{d}}(\tau) \ ,\ee

\be \label{int2} I^{(2)}_{abcd}=-i \int_{\tau_0}^0 d\tau \ a(\tau) \  \bar{\delta \phi}^*_{k_{a}}(\tau)\bar{\delta \phi}^*_{k_{b}}(\tau)\bar{\delta \phi}'^*_{k_{c}}(\tau)\bar{\delta \phi}^*_{k_{d}}(\tau) \ , \ee

\be \label{int3}  I^{(3)}_{abcd}=-i  \int_{\tau_0}^0 d\tau \ a(\tau)  \ \bar{\delta \phi}'^*_{k_{a}}(\tau)\bar{\delta \phi}'^*_{k_{b}}(\tau)\bar{\delta \phi}'^*_{k_{c}}(\tau)\bar{\delta \phi}^*_{k_{d}}(\tau) \ ,\ee
and
\be  \vec{k}_{ab}=\vec{k}_a+\vec{k}_b, \ \ \ \  k_{ab}=|\vec{k}_a+\vec{k}_b|, \ \ \ \vec{Z}_{ab}=\sigma_{ab} \, \vec{k}_a+\sigma_{ba} \, \vec{k}_b \ \ \ {\rm (no \ summation)} \ , \ \ \  \sigma_{ab}=\vec{k}_a\cdot \vec{k}_b+k_b^2 \ . \ee\\
The 
graviton exchange
contribution is given by 
\bea \label{GE} & &\langle \hat{\delta \phi}_{\vec{k}_1}\hat{\delta \phi}_{\vec{k}_2}\hat{\delta \phi}_{\vec{k}_3}\hat{\delta \phi}_{\vec{k}_4}\rangle_{GE} =(2\pi)^3 \delta(\sum_a \vec{k}_a) \ 4 \\ \nonumber & & \times \Big[ k_1^2 k_3^2 [1-(\hat{k}_1 \cdot \hat{k}_{12})^2][1-(\hat{k}_3 \cdot \hat{k}_{12})^2] \cos{(2 \chi_{12,34})} \cdot  (I_{1234}(0)+I_{3412}(0)) \\ \nonumber & & + (2\rightarrow 3,3\rightarrow 4)+ (2\rightarrow 4,3\rightarrow 2,4\rightarrow 3)\Big] \ , \eea
where $\hat{k}\equiv \vec{k}/k$, and
\bea \label{int4} I_{abcd}(\tau_*)=\int_{\tau_0}^{\tau_*} d\tau \int_{\tau_0}^{\tau} d\tau' \ a^2(\tau) \ a^2(\tau') \ {\rm Im} [\bar{\delta \phi}_{k_a}(\tau_*)\bar{\delta \phi}_{k_b}(\tau_*) \bar{\delta \phi}^*_{k_a}(\tau)\bar{\delta \phi}^*_{k_b}(\tau)] \\ \nonumber \times \ {\rm Im} [\bar{\delta \phi}_{k_c}(\tau_*)\bar{\delta \phi}_{k_d}(\tau_*) \bar{\delta \phi}^*_{k_c}(\tau') \bar{\delta \phi}^*_{k_d}(\tau')\bar{\gamma}_{k_{ab}}(\tau)\bar{\gamma}^*_{k_{ab}}(\tau')] \ .\eea
In the previous expression $\chi_{ab,cd}$ is the angle between the projections of $\vec{k}_a$ and $\vec{k}_c$ on the plane orthogonal to $\vec{k}_{ab}$ (see \cite{seery-sloth-vernizzi} for further details),  and $\gamma_{k}(\tau)$ are the mode functions describing tensor metric perturbations.
We recall that $\bar{\delta \phi}_{k}(\tau)=\alpha_k \ {\delta \phi}_{k}(\tau)+\beta_k \ {\delta \phi}_{k}^*(\tau)$, where ${\delta \phi}_{k}(\tau)$ are the 
Bunch-Davies 
modes in de Sitter space, that can be obtained from equation (\ref{modes})  together with the relation $\zeta_k=(H/\dot{\phi}_0) \delta \phi_k$, and $\tau_0$ is the time at which the initial state is specified at the onset of inflation. The integrals (\ref{int1},\ref{int2},\ref{int3},\ref{int4})  above, although very lengthy, can be computed straightforwardly, and the full expression for the trispectrum can be obtained. The detailed expression is not of much interest for the purposes of the present paper, and we do not  write it explicitly here. 

The relevant novel feature in the previous computation is that, as already happens for the bispectrum, the presence of initial quanta introduces {\em new contributions} to the correlation functions as a consequence of perturbative interactions among the quanta resulting from the stimulated creation process. These new contributions are characterized by being proportional to factors of the type $1/(\pm k_1\pm k_2\pm k_3\pm k_4)^n$, where the value of $n$ depends on the correlation function under consideration.\footnote{The presence of factors of the type $1/(\pm k_1\pm k_2\pm k_3\pm k_4)^n$ do not introduce divergencies for momentum configurations for which any of those denominators vanish. The complete expressions for the trispectrum and bispectrum are finite for all momentum configurations.}  These new factors, absent in the computation that uses an initial vacuum state, can be large for some specific momentum configurations, and are responsible for the enhancement that we find below in this section.

We will focus now on the squeezed configuration in which one of the four momenta 
in the trispectrum 
is much smaller than the rest, for instance, $k_3\ll k_1,k_2,k_4$. In the squeezed limit, the quadrilateral formed by the four momenta, as a consequence of the presence of the delta function in (\ref{trispectrum}), approaches a planar triangle. Among all possible squeezed configurations, we find of special relevance the one in which the three momenta $\vec{k}_1$, $\vec{k}_2$, $\vec{k}_4$ degenerate to a flattened triangle, as for instance the configuration shown in Fig.1 for which $\vec{k}_1\approx \vec{k}_2\approx \vec{k}_4/2$. It is worth 
mentioning that our search has 
not been exhaustive, and other configuration with similar enhancements may exist.

\begin{figure}[htbp] \label{figure}
\begin{center}
\includegraphics[width=6cm]{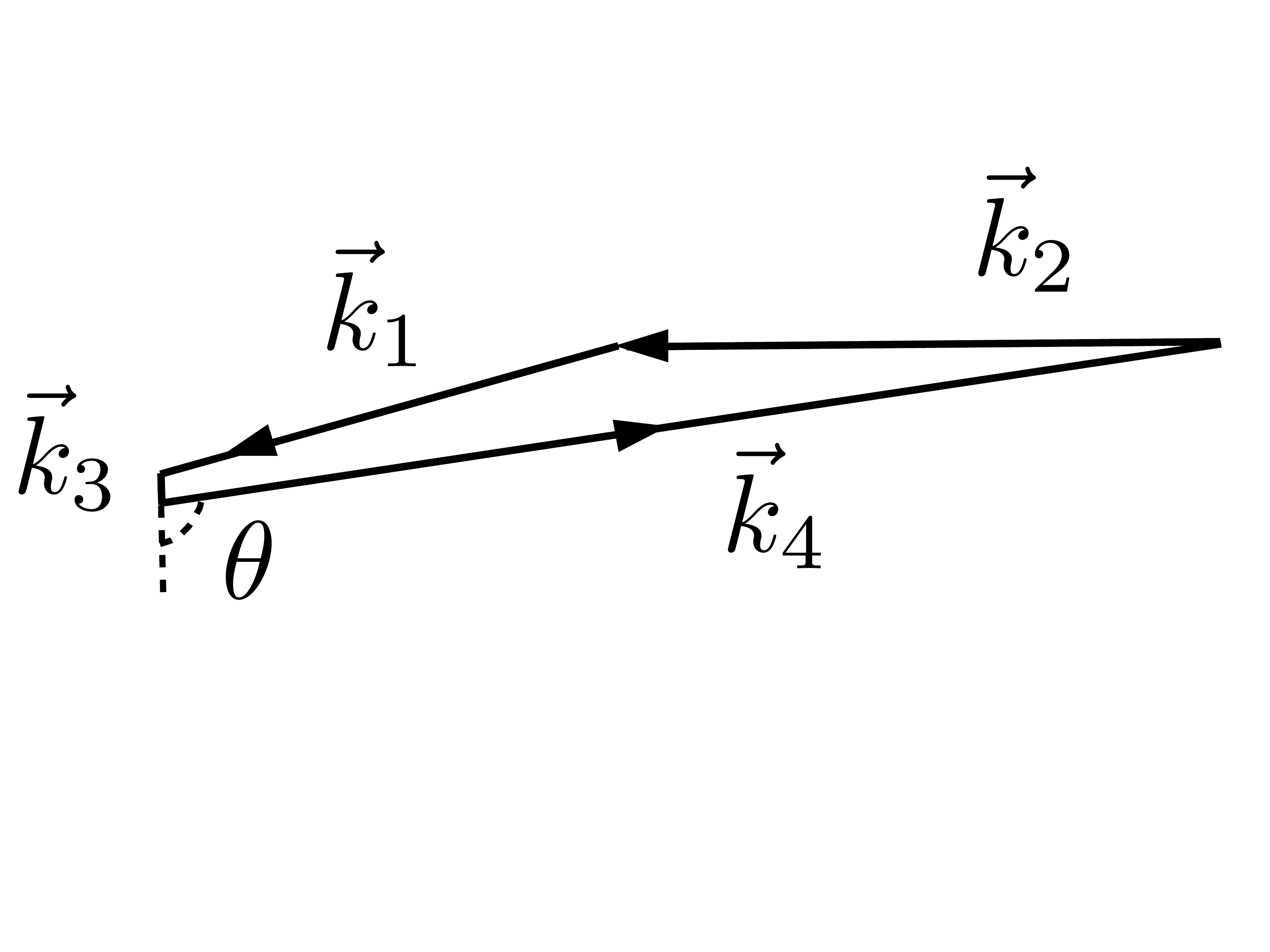}
\end{center}
\caption{Squeezed configuration. The larger three momenta generate a flattened triangle.}
\end{figure}

For this configuration, it is easy to show that, as a consequence of the geometrical factors shown in the second line of equation (\ref{GE}), the 
contact interaction
contribution largely dominates over the 
graviton exchange 
contribution. It is then straightforward to compute the contact interaction contribution using equation (\ref{CI}). Explicit computations show that

\be \langle \hat\zeta_{\vec{k}_1}\hat\zeta_{\vec{k}_2}\hat\zeta_{\vec{k}_3}\hat\zeta_{\vec{k}_4}\rangle=(2 \pi)^3 \delta(\sum_a \vec{k}_a) \frac{H^6}{4 \epsilon^2 M_P^6} \frac{\prod_a(\alpha_{k_a}+\beta_{k_a})}{\prod_a (2 k_a^3)}  \,  {\cal{M}}(\vec{k}_1,\vec{k}_2,\vec{k}_3,\vec{k}_4) \, +\, {\rm c.c.}\ , \ee 
where
\be \label{M} {\cal{M}}(\vec{k}_1,\vec{k}_2,\vec{k}_3,\vec{k}_4)=\frac{k_4^5}{k_3^2} \  h_{\alpha,\beta} \ , \ee
where the function $h_{\alpha,\beta}$ depends on the details of the initial state, and is given by

\be h_{\alpha,\beta}=\frac{3}{2} (\beta^*_{k_1}\beta^*_{k_2}\alpha^*_{k_4}+\alpha^*_{k_1}\alpha^*_{k_2}\beta^*_{k_4}) (\alpha^*_{k_3 }-\beta^*_{k_3}) \cos{\theta}\ , \ee
where $\theta$ is the angle between $\vec{k}_3$ and $\vec{k}_4$. Introducing equation (\ref{M}) in (\ref{CI}), we get the result for the trispectrum in the squeezed configuration shown in Fig. 1

\be \label{result} T_{\zeta}(\vec{k}_1,\vec{k}_2,\vec{k}_3,\vec{k}_4)= \ \hat g_{NL} \ P_{\zeta}(k_1)  P_{\zeta}(k_3) P_{\zeta}(k_4) \ee

\be \label{tau1} \hat g_{NL}= 32 \, \epsilon \left(\frac{k_1}{k_3}\right)^{2} f_{\alpha,\beta} ,\ee

where
\be \label{f} f_{\alpha,\beta}=2 \ {\rm Re}  \left[ \frac{h_{\alpha,\beta}\  \cdot\prod_{a} (\alpha_{k_a}+\beta_{k_a})}{\prod_{a={1,3,4} }|\alpha_{k_a}+\beta_{k_a}|^2 }\right] \ , \  \  \  \  P_{\zeta}(k)=|\bar{\zeta}_k|^2=\frac{1}{2 \epsilon M_P^2} \frac{H^2}{2 k^3} |\alpha_{k}+\beta_{k}|^2\ .\ee  

The parameter $\hat g_{NL}$ introduced above for the squeezed configuration is related to the conventional parameters $\tau_{NL}$ and $g_{NL}$, defined by the so called local form of the trispectrum\footnote{This form of the trispectrum arises from the phenomenologically motivated form of the curvature perturbation in position space 
$\hat\zeta=\hat\zeta_g+ 1/2 \, (\tau_{NL})^{1/2}\, (\hat\zeta_g^2-\langle \hat\zeta_g^2\rangle )+9/25 \, g_{NL}\, \hat\zeta_g^3$, where $\hat\zeta_g$ is a  Gaussian field.}
\bea T^{\rm local}_{\zeta}(\vec{k}_1,\vec{k}_2,\vec{k}_3,\vec{k}_4)&=&  \tau_{NL} \, [P_{\zeta}(k_1) P_{\zeta}(k_2)(P_{\zeta}(k_{13}) + P_{\zeta}(k_{14})) + 11 \ {\rm perm.} ] \nonumber \\ &+& \frac{54}{25}\, g_{NL}\, [P_{\zeta}(k_1) P_{\zeta}(k_2)P_{\zeta}(k_3) + 3 \ {\rm perm.}] \ , \eea
by the relation
\be \hat g_{NL}= 20 \, \tau_{NL} + \frac{540}{25} \, g_{NL} \ . \ee
Expressions (\ref{result}) and (\ref{tau1}) constitute the main result of the present paper.   Note that in the case that the average number of initial quanta is considerably smaller than 1, $|\beta_k|^2  \ll 1$, additional terms not explicitly shown in  (\ref{result}) and (\ref{tau1})  become relevant  and the vacuum result is smoothly recovered.

\section{Discussion and conclusions} \label{discussion and conclusions}
We have calculated the primordial trispectrum for curvature perturbations in inflationary models with canonical kinetic terms, when the initial quantum state is different from  the Bunch-Davies vacuum.  
The effects of the quanta initially present do not dilute with the exponential inflationary expansion, and survives to the end of inflation and beyond, as a consequence of the process of stimulated creation of quanta  in the expanding background. The presence of initial quanta introduces {\em new contributions} to the correlation functions.  
These contributions result
from perturbative 
interactions 
among the quanta produced by the stimulated creation process. 

We have focused our computations on the  squeezed momentum configuration, in which one of the four momenta is much smaller than the others. The squeezed limit is interesting due to the existence of consistency relations that constrain the magnitude of the $n$-point correlation functions, for $n>2$, generically known as the Maldacena consistency relations \cite{maldacena, creminelli-zaldarriaga,chen}. If the bispectrum and trispectrum are parametrized as in equations (\ref{bi}) and (\ref{result}), respectively, the consistency relations imply that $f_{NL}=\mathcal{O}(\epsilon)$ for the bispectrum and $\hat g_{NL}=\mathcal{O}(\epsilon^2)$ for the trispectrum, with the only assumption being that the inflaton field is the only dynamical field. 
Those values 
of the bispectrum and trispectrum are well below the observational threshold of the next generation of detectors, since $\epsilon=\mathcal{O}(10^{-2})$. Therefore, the consistency 
relations imply 
that any detection of a primordial $f_{NL}$ or $\hat g_{NL}$ will strongly disfavor single-field inflation. 

However, although the consistency relations may be true under very generic assumptions in the {\em exact} limit in which one of the momenta goes to zero, $k_a\to 0$, it is possible that important corrections appear when restricted to the range of momenta accessible in our finite observable universe. It was shown in \cite{agullo-parker} that this is the case for $f_{NL}$ when the initial state for perturbations at the onset of inflation departs from the vacuum. There, it was shown that, for  
a large class of initial states,
the $f_{NL}$ parameter acquires an extra factor of the order $k_1/k_3\gg1$, as compare to the vacuum prediction.  
That enhancement factor can be as large as several hundred for the range of $k's$ accessible in forthcoming observations, which may cause the bispectrum in the squeezed momentum configuration to fall within the observational sensitivity of the PLANCK satellite \cite{ganc}.

In the present paper we have focused our attention on the trispectrum. We have considered an initial state given by a Bogolubov transformation of the Bunch-Davies vacuum, although our main conclusions apply also to more generic states. 
Among all possible squeezed configurations for the trispectrum, we find of special relevance the configurations for which the larger momenta form a flattened triangle, as for instance the one shown in Fig.1 for which $k_1\approx k_2\approx k_4/2\gg k_3$.  For that configuration the trispectrum acquires the local form (\ref{result}) with a momentum dependent amplitude given by 

\be \label{tau} \hat g_{NL}= 32 \, \epsilon \left(\frac{k_1}{k_3}\right)^{2} f_{\alpha,\beta} \, ,\ee 
where $f_{\alpha,\beta}$, written in equation (\ref{f}), is a function that contains the information about the initial state.  Because $f_{\alpha,\beta}$ is a ratio of Bogolubov coefficients and, as mentioned at the end of section \ref{quantum state}, the values of the $\alpha_k$ and $\beta_k$ coefficients are constrained not to vary arbitrarily fast with $k$, its value is not very sensitive to the peculiarities of the initial state. When the average number of initial quanta in the observable modes is of order one or greater, $f_{\alpha,\beta}$ is generally of order one. 
Therefore, the
above value for $\hat g_{NL}$ is enhanced by a factor $\mathcal{O}(\epsilon^{-1} (k_1/k_3)^2)$ as 
compared
to the prediction of the consistency relation. The range of $k$'s for which we have some confidence about the measurements of the temperature fluctuations of the CMB (i.e., for which uncertainties coming from cosmic variance, Sunyaev-Zeldovich effect, etc., can be neglected) is approximately 
several hundred. 
Therefore, in the squeezed limit we find an enhancement of the $\hat g_{NL}$ parameter that can be as large as $\mathcal{O}(10^6)$, 
compared
to the 
prediction
of the consistency relation for the trispectrum. Equation (\ref{tau}) shows 
that $\hat g_{NL}$
can be as large as $\hat g_{NL}=\mathcal{O}(10^3)$. This value is  below the current experimental bound \cite{smidt, kogo-komatsu}. 
However, it may reach the sensitivity of the next generation of observations.

On the other hand, large scale structure surveys and halo bias measurements may provide further constraints on the non-Gaussianities reported in this paper, as those measurements are particularly sensitive to non-Gaussianities arising from the squeezed 
momentum 
configurations \cite{shandera}. The scale-dependent amplitude of the bispectrum and trispectrum arising from a non-vacuum initial state 
\be f_{NL}\propto \epsilon \,  \frac{k_1}{k_3} \ , \hspace{2cm} \hat g_{NL}\propto \epsilon \left(\frac{k_1}{k_3}\right)^2\ , \ee
provides a clear signature to distinguish the effects of the initial conditions for perturbations at the onset of inflation from other known sources. 

 The large local-type bispectrum and trispectrum arising from a non-vacuum initial state is not exclusive of inflationary models with canonical kinetic term. In fact, an enhancement in the trispectrum in the squeezed limit is also present in other models of inflation with more general lagrangians. Such enhancement can be even larger than the one found in this paper for canonical kinetic terms, and its details would require further analysis.

The results given in this paper are of relevance to 
the interpretation of 
the forthcoming observations of non-Gaussianities, as an observation of large local-type non-Gaussianities 
would 
not necessarily 
rule out 
single-field inflation, as widely believed. On the contrary, such observations could be interpreted as a consequence of having an initial state for inflation that contains initial perturbations 
that have effects observable today as a result of the stimulated creation of quanta during inflation. 
Such an interpretation could be tested through the observational signatures we discussed above.

\acknowledgments  I.A. thanks S. Shandera for helpful discussions and comments. We thank J. Ganc and E. Komatsu for useful comments on the manuscript. This work has been partially supported by the Spanish grants FIS2008-06078-C03-02, FIS2010-09399-E, the Consolider Program CPAN (CSD2007-00042), and NSF grants PHY-0503366 and PHY0854743 and Eberly research funds of Penn State University. The Institute for Gravitation and the Cosmos is supported by the Eberly College of Science and the Office of the Senior Vice President for Research at the Pennsylvania State University.  J. N-S thanks MEC for a sabbatical grant and the Physics Department of the University of Wiscosin-Milwaukee for their kind hospitality. \\

\end{document}